%% file: acl_latex.tex
\documentclass[11pt]{article}

\usepackage[preprint]{acl}

\usepackage{times}
\usepackage{latexsym}

\usepackage[T1]{fontenc}

\usepackage[utf8]{inputenc}

\usepackage{microtype}

\usepackage{inconsolata}

\usepackage{graphicx}

\usepackage{colortbl}  
\usepackage{xcolor}  
\usepackage{amsmath}
\usepackage[utf8]{inputenc}
\usepackage[most]{tcolorbox}
\usepackage{pgfplots}
\usepackage{enumitem}
\usepackage[table]{xcolor}
\usepackage{latexsym}

\usepackage{pifont}
\usepackage{booktabs}
\usepackage{amssymb}
\usepackage{times}
\usepackage{adjustbox}
\usepackage{multirow}
\usepackage{caption}
\usepackage{subcaption}
\usepackage{float}
\definecolor{nBlue}{RGB}{0,165,249}
\definecolor{nGreen}{rgb}{0, 0.5, 0.2}
\definecolor{nRed}{rgb}{0.8, 0.1, 0.2}
\definecolor{mGreen}{rgb}{0.3, 0.65, 0.4}
\usepackage[ruled, vlined, noend, linesnumbered]{algorithm2e} 
\SetKwFor{Upon}{upon}{do}{end upon}
\usepackage{bbm}
\title{Agent Harness Distillation: Inference-Time Harness Extraction and Exploitation in Autonomous Multi-Agent Systems}

\author{
\textbf{Yu Cui}\textsuperscript{1}  \quad
\textbf{Wuli Yang}\textsuperscript{1}  \quad
\textbf{Yirui Shi}\textsuperscript{1}  \quad
\textbf{Junhao Xia}\textsuperscript{1}  \\
\textbf{Hui Jiang}\textsuperscript{1,2}  \quad
\textbf{Lei Gao}\textsuperscript{1}\thanks{Corresponding authors.} \quad
\textbf{Chenfu Bao}\textsuperscript{1,2}\footnotemark[1]
\\ 
\textsuperscript{1}Baidu Inc. \quad
\textsuperscript{2}Tsinghua University \\
\texttt{\{cuiyu08, yangwuli, gaolei01, baochenfu\}@baidu.com} 
}

\pgfplotsset{compat=1.18}
\begin{document}
\maketitle
\begin{abstract}
Autonomous multi-agent systems (AMAS) built on large language models (LLMs), such as Hermes, increasingly rely on inference-time harnesses to coordinate reasoning and action. Constructing these harnesses requires substantial engineering effort and computational resources, as they are iteratively optimized over a combinatorial search space while co-evolving with the underlying LLM. Inference-time harnesses therefore constitute valuable intellectual property (IP). Although prior work has investigated IP leakage in static multi-agent systems with pre-configured architectures, it remains unclear whether similar risks arise in AMAS, where harness behavior emerges dynamically during inference. To address this gap, we introduce Agent Harness Distillation (AHD), a framework for studying the security risks arising from inference-time harness extraction in AMAS. We formalize harness extraction as a new security problem and develop an evaluation framework for quantifying such risks. AHD extracts inference-time harness capabilities from a target agent through black-box interactions and consists of two stages. In the pre-distillation stage, AHD infers inference-time harness behaviors from the responses of the target agent and constructs an initial harness. In the post-distillation stage, AHD iteratively refines the initial harness to align with the behavioral patterns of the target agent. Experiments on real-world AMAS across multiple backbone LLMs demonstrate the effectiveness of AHD and reveal substantial IP leakage risks. We further propose a deception-based defense that reduces harness extraction effectiveness while preserving the utility of the protected agent. Our findings uncover a previously underexplored security threat to AMAS.
\end{abstract}

\section{Introduction}
The rapid development of large language models (LLMs) has accelerated the deployment of LLM agents in real-world software development \citep{wang2026shadows, qian-etal-2024-chatdev}. In particular, Autonomous Multi-Agent Systems (AMAS), such as Claude Code and Hermes, organize multiple agents, tools, and planning modules into automated workflows, substantially improving developer productivity and lowering the barrier to complex programming tasks \citep{liu2026clawkeeper}. For each user task, these systems dynamically instantiate an inference-time harness \citep{zhao2026training}, which specializes a static infrastructure scaffold into a task-specific workflow. By orchestrating planning, tool use, and inter-agent collaboration, the harness substantially enhances the problem-solving capabilities of the backbone LLM.

However, designing an effective harness requires substantial engineering effort, manual experimentation, and iterative co-evolution with the underlying LLM. This process is further complicated by the large search space induced by choices over planning, tool orchestration, subagent roles, and inter-agent coordination \citep{zhou2026multiagent, xu2026adapting, chen2026harnessforge, lee2026recursive}. Consequently, inference-time harnesses represent valuable intellectual property (IP). Meanwhile, they introduce a new security concern: an adversary interacting with an AMAS may recover information about the inference-time harness, thereby extracting reusable system knowledge that leads to IP leakage.

Prior work has shown that components of static multi-agent systems can be extracted through interaction, leading to potential IP leakage \citep{wang2026masleak}. However, it remains unclear whether similar threats extend to AMAS. Unlike static multi-agent systems, task-dependent workflows in AMAS are dynamically instantiated at inference time rather than predefined.

\noindent
\textbf{Challenges}. A systematic study of inference-time harness leakage in AMAS faces three key challenges. First, inference-time harnesses are dynamically instantiated during execution, making the leaked information difficult to define and formally characterize. Second, the severity of such leakage is difficult to measure. An inference-time harness comprises both explicit structural components and implicit execution strategies, dimensions that existing metrics do not adequately capture. Third, the practical security impact remains unclear. Extracted harness information may not translate into actionable capabilities or effective exploitation. To address these challenges, we investigate the following research questions:

\begin{itemize}[left=0pt, itemsep=0pt]
\item \textbf{RQ1}: How can inference-time harness leakage in AMAS be formally defined and quantified?
\item \textbf{RQ2}: Can inference-time harness information be extracted through black-box interactions, and what practical value does such leakage provide?
\item \textbf{RQ3}: How can inference-time harnesses be protected against extraction attacks?
\end{itemize}

To answer RQ1, we formalize inference-time harness leakage in AMAS by introducing a unified abstraction that captures dynamically instantiated orchestration structures and execution strategies. We further develop an evaluation framework with four core metrics to quantify different dimensions of harness leakage.
To answer RQ2, we investigate inference-time harness extraction attacks and their downstream exploitation in AMAS. Specifically, we propose \textit{Agent Harness Distillation} (AHD), which extracts harness information from a target agent and transfers the corresponding orchestration capabilities to another agent. AHD consists of two stages. In the pre-distillation stage, AHD infers harness behaviors from observable agent responses and constructs an initial harness. In the post-distillation stage, AHD iteratively refines the initial harness to better match the target agent's behavioral patterns.
To answer RQ3, we introduce a deception-based defense that provides plausible but misleading harness information in response to extraction attempts, reducing extraction fidelity while preserving the utility of the protected agent. Our contributions are as follows:

\begin{itemize}[left=0pt, itemsep=0pt]
\item We formalize inference-time harness leakage as a new security problem 
    in AMAS and develop an evaluation framework to quantify the associated 
    leakage risks.

\item We propose AHD, a two-stage framework that extracts inference-time harness information from target agents through black-box interactions and transfers the corresponding orchestration capabilities. Experiments on real-world AMAS across multiple backbone LLMs demonstrate the effectiveness of AHD and reveal substantial IP leakage risks.

\item We design a deception-based defense that provides misleading harness information to extraction attempts, reducing extraction fidelity while preserving the utility of the protected agent.
\end{itemize}

\begin{figure*}[t]
    \centering
    \includegraphics[width=1\linewidth]{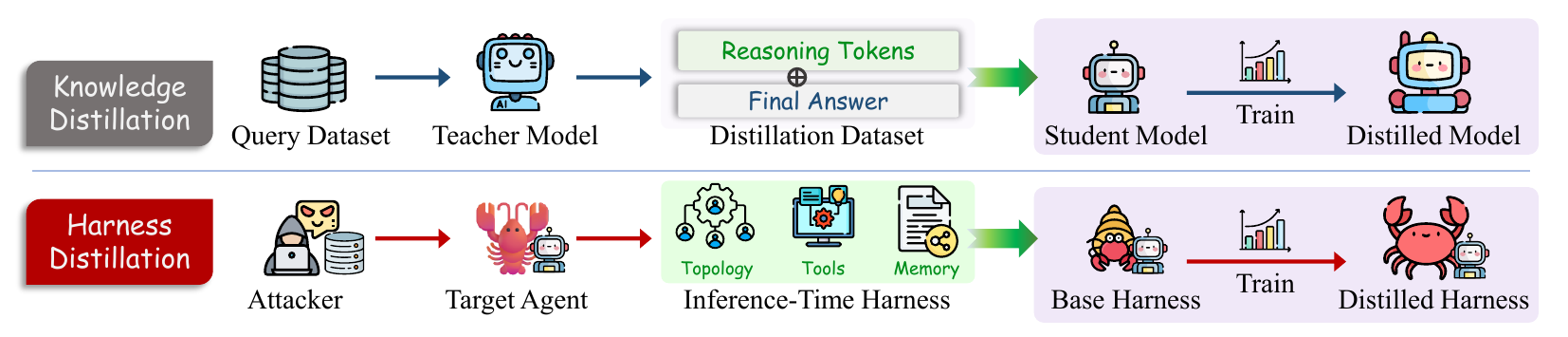}
    \caption{Comparison of the pipelines for our agent harness distillation and model knowledge distillation.}
    \label{fig:overview}   
\end{figure*}

\input{latex/Related_Work}
\input{latex/Methodology}

\input{latex/Experiment}

\input{latex/Discussion}

\section{Conclusion}
In this paper, we investigate inference-time harness leakage in AMAS and formulate harness extraction as a new security problem. We propose Agent Harness Distillation (AHD), a two-stage black-box framework for extracting harness behaviors from target agents and transferring their orchestration capabilities. Extensive experiments across real-world AMAS with multiple backbone LLMs show that AHD can effectively extract transferable harness capabilities, exposing substantial IP leakage risks. We further present a deception-based defense that mitigates harness extraction while maintaining agent utility. Our work reveals a previously underexplored attack surface in AMAS. Beyond this security implication, our study also provides insights into harness self-evolution by showing the potential of extracting and transferring inference-time harness behaviors.

\section{Ethical Considerations}
This paper introduces an attack that extracts proprietary information about harness structures in MASs. Our goal is to improve the security of LLM agents by revealing a previously unexplored attack surface and developing rigorous red-teaming methods. We also analyze the associated risks and present corresponding defense strategies. The techniques described in this paper are intended solely for scientific research. AI assistants are used to polish the writing.


\bibliography{acl_latex}
\onecolumn
\newpage

\appendix

\input{latex/Appendix}

\end{document}

%% file: latex/Related_Work.tex
\section{Related Work and Background}
\textbf{IP Leakage in MAS}.
Existing work on IP leakage in MAS shows that communication topologies, often designed with substantial effort, can be partially recovered through external interactions \citep{wang2026masleak}. Such topologies are not only valuable system-level IP, but also useful priors for follow-up attacks. Once the communication relations among agents are known, many attacks that assume a black-box setting \citep{amayuelas-etal-2024-multiagent, li2026flowsteer} can be turned into cheaper and more targeted gray-box attacks, enlarging the attack surface \citep{wu-etal-2026-cia, an-etal-2026-aciarena}. However, prior work mainly targets traditional MAS with static topologies, and the attack objects are usually predefined topologies such as linear, star, tree, complete, and random. In these systems, the topology is fixed before task execution, so the inference-time topology remains stable across queries. By contrast, practical AMAS instantiate an inference-time harness dynamically during inference. This harness includes not only topology, but also context assembly, tool
interaction, and memory management \citep{huang2026memoharness}. It exists only for a single task run and is discarded afterward. Its transient and dynamic nature makes extraction and recovery much harder than static topology extraction. Moreover, existing methods often rely on privileged access to intermediate outputs from subagents, an assumption that does not hold in real-world AMAS deployments that expose only a unified user interface. To fill this gap, we extend the target from static MAS topologies to AMAS inference-time harnesses, and study how to extract and recover them under a strict black-box setting.

\noindent
\textbf{Knowledge Distillation and Harness Self-Evolution}.
Knowledge distillation \citep{hayder2026dtokd, he2025dakd, panfilov2026stealing} is a widely adopted paradigm for transferring knowledge from a high-capacity teacher to a compact student model, enabling effective model compression. However, distillation for LLMs has also raised concerns over capability replication from closed-source LLMs and commercial IP protection \citep{zhao2025can}. Meanwhile, for LLM agents, harnesses play a critical role in governing agent behavior and system capabilities. They can further self-evolve through training or iterative optimization \citep{huang2026memoharness, chen2026harnessforge, wang2026handbook, chen2026harnessx, jiang2026self}. This process, however, demands high computational resources and long optimization cycles due to cascaded evaluation rounds. It also requires access to internal model states \citep{zhou2026multiagent}. Existing research has not explored whether system-level harnesses can be distilled like model knowledge. To bridge this gap, we introduce \textit{Agent Harness Distillation}, a new paradigm that extends distillation from model parameters to inference-time harnesses. We investigate whether recovered harnesses can be transferred to weaker agents to reproduce the capabilities of source agents, revealing a new form of system-level IP leakage.

%% file: latex/Methodology.tex
\section{Preliminary Study}
\subsection{Motivation and Research Settings}
Some AMAS have begun to encrypt prompts and task messages exchanged among agents. For example, Codex MultiAgentV2\footnote{https://github.com/openai/codex/issues/28058} records task information delegated from the primary agent to subagents as ciphertext. As a result, user developers cannot directly obtain the specific delegation content from local execution records. This encryption mechanism reduces the risk of IP leakage for agent providers. However, existing protection mainly focuses on inter-agent messages. Potential leakage of the inference-time harness through responses remains underexplored. Future AMAS may further restrict access to internal system logs, for example, by encrypting these logs \citep{carolina2025eal}. In this setting, black-box responses will become a primary channel through which external observers infer internal orchestration structures. Motivated by this observation, we study a harness extraction attack under a strict black-box setting. The attacker can only submit queries through a standard interface and observe the responses. This setting demonstrates a practical security threat in real-world deployments. 

\subsection{Problem Formulation}
\label{sec:problem_formulation}
\input{latex/form}
\input{latex/distill}

\input{latex/defense}
\input{latex/metrics}

%% file: latex/form.tex
\textbf{System Model}.
We consider a black-box AMAS $\mathcal{A}=(\theta,\mathcal{H})$, where \(\theta\) denotes the underlying foundation LLM and
\(\mathcal{H}\) denotes the static harness. For a query \(x\in\mathcal{X}\), the agent $\mathcal{A}$ first constructs a task-specific inference-time harness
$
h_x=\mathcal{I}_{\mathcal{H}}(x),
$
where \(\mathcal{I}_{\mathcal{H}}\) is the harness instantiation function. The instantiated harness \(h_x\) specifies the execution strategy, including agent coordination, context assembly, tool interaction, and memory management. The final response is generated by executing the foundation LLM under the instantiated harness:
\[
y=\operatorname{Exec}(\theta,h_x,x).
\]

Different from persistent model parameters or stored configurations, $h_x$ is an ephemeral inference-time state. It only exists during the execution process and is discarded after producing the final response. Therefore, although the harness determines the agent's behavior, it is not directly accessible after inference.

\noindent
\textbf{Threat Model}.
We consider a black-box interaction setting, where an attacker can query the agent system $\mathcal{A}$ but cannot access its internal execution states. 
Following real-world LLM agent deployments (e.g., Claude Code), we assume the attacker can locally deploy the agent and select the underlying backbone model, but only has black-box interaction access without observing internal execution states. From a practical perspective, the attacker typically accesses $\mathcal{A}$ as a regular user and can freely configure the underlying backbone model. However, the attacker has no access to the harness $\mathcal{H}$, the instantiated inference-time harness $h_x$, or the internal execution logs $\mathcal{L}$. We assume that the system protects its internal logs through access control, encryption, or other isolation mechanisms. During inference, $h_x$ is recorded in the internal logs of $\mathcal{A}$:
$
h_x\rightarrow \mathcal{L}.
$
However, neither \(h_x\) nor \(\mathcal{L}\) is directly observable by the attacker. The only observable signal is the final response $y$. Therefore, the attacker can only infer the hidden $h_x$ through designed queries and analysis of the corresponding responses.

\noindent
\textbf{Harness Extraction Attack}.
Given a task $x$, the attacker constructs a query
$
x'=x \circ d,
$
where $d$ is an injected probe designed to induce the agent $\mathcal{A}$ to reveal information about its $h_{x'}$. The resulting response is
$
y'=\operatorname{Exec}(\theta,h_{x'},x').
$
An extraction function $E$ is applied to the observable response to obtain an estimated harness:
\[
\hat h_{x'}=E(y').
\]
The attacker aims to recover the original inference-time harness $h_x$ corresponding to the clean query $x$, rather than the potentially modified harness induced by the injected query $x'$. To formalize this distinction, let
$
h_x=\mathcal{I}_{\mathcal{H}}(x)
$
denote the clean harness and
$
h_{x'}=\mathcal{I}_{\mathcal{H}}(x')
$
denote the harness under attack. The injected probe may alter the agent's execution behavior, causing the instantiated harness under attack to deviate from that induced by the clean query. We characterize this deviation using a distance function
$D(\cdot,\cdot)$ between two instantiated harnesses. Specifically, the attack perturbation is defined as
\[
\epsilon_A=D(h_x,h_{x'}),
\]
which measures the execution discrepancy introduced by the injected probe. Since the attacker can only infer the harness from the observable response, the recovered harness inevitably incurs an additional extraction error:
\[
\epsilon_E=D(h_{x'},\hat h),
\]
where $\hat h$ denotes the recovered harness. The overall recovery error is therefore decomposed into two components:
\[
\epsilon=\epsilon_A+\epsilon_E.
\]
\noindent
Accordingly, the objective of harness extraction is to jointly optimize the injected probe $d$ and the extraction algorithm $E$ by minimizing the overall recovery error:
\[
(d^*,E^*)
=
\arg\min_{d,E}
\epsilon
=
\arg\min_{d,E}
(\epsilon_A+\epsilon_E).
\]
\noindent
An ideal extraction attack satisfies
$
\epsilon_A\rightarrow0,
\epsilon_E\rightarrow0,
$
which implies
$
\hat h\approx h_x.
$

\noindent
\textbf{Learning-Theoretic View}.
From a learning-theoretic perspective, harness extraction can be viewed as a new inverse problem \citep{authors2024inverse}. Unlike conventional supervised learning, which aims to learn an input-output mapping, the execution process of an AMAS is only partially observed. Specifically, the inference process can be formulated as
$
y=\operatorname{Exec}(\theta,h,x),
$
where the backbone model $\theta$, the input query $x$, and the final response $y$ are observable, while the instantiated inference-time harness $h$ remains hidden. Consequently, instead of learning a predictive function
\[
f: x \overset{h}{\longrightarrow} y,
\]
the attacker seeks to recover $h$ from observable interactions:
\[
(\{\theta,x\},y)\longrightarrow h.
\]
Once instantiated harnesses are recovered from sufficiently diverse interactions, they can be aggregated to estimate the underlying static harness $\hat{\mathcal H}$, which approximates the original $\mathcal{H}$ of $\mathcal{A}$. The reconstructed system therefore satisfies
\[
\operatorname{Exec}(\theta,\hat{\mathcal H},x)
\approx
\operatorname{Exec}(\theta,\mathcal H,x).
\]

Unlike conventional knowledge distillation, harness extraction targets latent execution mechanisms rather than model knowledge. This perspective naturally leads to agent harness distillation, which transfers recovered inference-time harnesses across backbone models to reproduce system-level capabilities without access to the original implementation, extending distillation beyond models to AMASs.

\noindent
\textbf{Harness Identifiability.}
Although harness extraction aims to recover the hidden harness $h$ of $\mathcal{A}$, exact recovery is not always possible since different harnesses may induce identical observable behaviors. Under a query distribution \(\mathcal{D}\), two harnesses $h_1$ and $h_2$ are behaviorally indistinguishable \citep{cheval2023csf} if

\[
\forall x\in\mathcal{D},\quad 
\mathrm{Exec}(\theta, h_1, x)=\mathrm{Exec}(\theta, h_2, x),
\]
\noindent
denoted as \(h_1 \sim h_2\). Therefore, black-box observations identify the target harness only up to an indistinguishability class. The extraction objective is thus to recover
$
\hat{h}\in[h],
$
where \([h]=\{h^{l}: h^{l}\sim h\}\) denotes the behavioral indistinguishability class of \(h\). A harness is identifiable up to behavioral indistinguishability if there exists an extractor \(E\) satisfying
\[
\Pr_{x\sim\mathcal{D}}
\left[
\min_{h^{l}\in[h]}D\bigl(h^{l}, E(y)\bigr)\leq\epsilon
\right]\geq1-\delta,
\]
\noindent
where $\epsilon$ denotes the tolerable recovery error, and \(\delta\) denotes the failure probability. 

%% file: latex/distill.tex
\section{Agent Harness Distillation}
\label{sec:method}
In this paper, we define AHD as follows: given any query $x \in \mathcal{X}$ and an agent $\mathcal{A}=(\theta,\mathcal{H})$, AHD aims to reconstruct an agent $\mathcal{B}=(\theta',\hat{\mathcal{H}})$ through black-box methods, such that the functional behavior of $\mathcal{H}$ can be transferred to $\hat{\mathcal{H}}$. Consequently, agents $\mathcal{A}$ and $\mathcal{B}$ achieve similar performance when executing queries $x$ over $\mathcal{X}$.
In this section, based on the formalization in Section \ref{sec:problem_formulation}, we present a harness distillation pipeline for recovering the static harness $\mathcal{H}$ of a target AMAS $\mathcal{A}$ from black-box interactions. The pipeline consists of two phases: \textit{Pre-Distillation} and \textit{Post-Distillation} (see Algorithm~\ref{alg}).

\subsection{Pre-Distillation}
\input{latex/predistill}

\subsection{Post-Distillation}
\input{latex/postdistill}

\input{latex/alg}

%% file: latex/predistill.tex
Given a task \(x_i\in\mathcal{D}\), the attacker first constructs an augmented query $x_i'=x_i\circ d$,
where $d$ is designed to elicit information about the target system's inference-time harness while preserving the original task objective. Specifically, each injected data sample $d$ consists of two components: a prefix and a suffix. The prefix is designed to induce the target agent to activate complex inference-time execution structures by encouraging the construction of auxiliary subagents and the invocation of external tools. To achieve this, the prefix introduces uncertainty regarding the reliability of the provided context, prompting the agent to verify potentially fabricated information through additional tool calls and subagent collaboration. This process naturally stimulates richer inference-time workflows, exposing latent harness components that are otherwise difficult to observe.
The suffix is designed to elicit the disclosure of these inference-time execution details after task completion. It guides the agent to inspect system logs, retrospectively analyze its execution trajectory, and summarize the underlying workflow. The induced harness deviation is measured as $\epsilon_A$, which characterizes the impact of the query perturbation on the instantiated harness. We refer to this attack as the Harness Extraction Attack (HEA). The attacker executes each query against the target agent $\mathcal{A}$ for $n$ independent runs, obtaining responses
$
\{y_{i,r}'\}_{r=1}^{n},
$
from which structured harness claims are extracted:
\[
\hat h_{i,r}=\operatorname{Extract}(y_{i,r}').
\]
Each claim \(\hat h_{i,r}\) describes the disclosed orchestration structure across multiple dimensions, such as agent roles, topology, tool usage, and coordination mechanisms.
To obtain a robust estimate of the latent harness, the collected claims are aggregated into a harness representation \(\hat{h}\). Specifically, for each structural dimension \(u\in\mathcal{U}\), we first collect the corresponding claims
$
\mathcal{Z}_u=
\{\hat h_{i,r}[u]\}_{i\in[N]}^{r\in[n]}
$.
We then perform two-stage denoising. First, per-task majority voting retains only claims consistently reproduced across independent runs, removing unstable self-reports. Second, cross-task frequency aggregation preserves only claims that appear consistently across different tasks, filtering task-specific artifacts while retaining system-level structural regularities. The resulting representation \(\hat{h}\) serves as the extracted harness template. During cross-task denoising, some legitimate $\hat h_{i,r}[u]$ values may be incorrectly discarded due to frequency-based filtering. However, such information can be recovered during the subsequent harness distillation process. Finally, $\hat{h}$ is instantiated into an executable MAS by a coding agent, which maps each recovered structural component to a corresponding execution strategy, producing the initial distilled harness $\mathcal{H}_0$ for subsequent adaptation.

%% file: latex/postdistill.tex
The harness \(\mathcal{H}_0\) obtained from pre-distillation serves as a basic underlying scaffold, which we refer to as the base harness. Given \(\mathcal{H}_0\) and the attacker-known backbone model \(\theta\), we construct the base agent as \(\mathcal{B}_0 = (\theta, \mathcal{H}_0)\).
This base agent also serves as the starting point for post-distillation. During post-distillation, we evaluate \(\mathcal{B}_0\) using the same inputs $x_i$ employed for harness extraction during pre-distillation, together with their corresponding prefixes in \(\mathcal{D}\). Because \(\mathcal{B}_0\) is instantiated by the attacker, its inference-time harness \(h_{x_i}^{\mathcal{B}_0}\) can be directly observed through execution logs or equivalent instrumentation. An additional coding agent \citep{xu2026adapting} then compares the observed harness \(h_{x_i}^{\mathcal{B}_0}\) against the previously recovered harness \(\hat{h}_{x_i}\). Based on the identified structural discrepancies, the coding agent proposes modifications to \(\mathcal{H}_0\), producing a candidate harness \(\mathcal{H}_i^{\mathrm{candidate}}\). Repeating this procedure over the samples in $\mathcal{D}$ yields an iterative process that we refer to as \textit{Loop Harness Alignment}. At iteration \(i\in\{1,\ldots,|\mathcal{D}|\}\), the current agent is defined as \(\mathcal{B}_{i-1} = (\theta,\mathcal{H}_{i-1})\). We execute \(\mathcal{B}_{i-1}\) on \(x_i\) and observe its inference-time harness \(h_{x_i}^{\mathcal{B}_{i-1}}\). A structural edit is then selected to reduce the discrepancy between the observed harness and the recovered target harness $\hat{h}_{x_i}$. At each iteration, the candidate agent is evaluated on a separate validation set \(\mathcal{V}\). Let \(e_{i-1}=\operatorname{Eval}(\mathcal{B}_{i-1},\mathcal{V})\) denote the validation performance of the current agent, and let \(e_i^{\mathrm{candidate}}=\operatorname{Eval}(\mathcal{B}_i^{\mathrm{candidate}},\mathcal{V})\) denote that of the candidate agent. The candidate edit is accepted only if it improves or preserves validation performance:
\[
\mathcal{H}_i
=
\begin{cases}
\mathcal{H}_i^{\mathrm{candidate}},
& e_i^{\mathrm{candidate}} \geq e_{i-1},\\[4pt]
\mathcal{H}_{i-1},
& \text{otherwise}.
\end{cases}
\]

To mitigate overfitting and preserve generalization, the validation set \(\mathcal{V}\) is disjoint from both the samples used to propose structural edits and the final test set. The procedure terminates when either a fixed iteration budget is exhausted or inference performance converges. The resulting adapted harness is denoted by \(\hat{\mathcal{H}}_B\).

%% file: latex/alg.tex
\begin{algorithm*}[t]
\caption{Two-Stage Harness Distillation}
\label{alg}
\DontPrintSemicolon
\SetKwInOut{Input}{Input}
\SetKwInOut{Output}{Output}
\SetKwComment{Comment}{$\triangleright$~}{}
\SetKwFunction{Exec}{Exec}
\SetKwFunction{Extract}{Extract}
\SetKwFunction{Instantiate}{Instantiate}
\SetKwFunction{Aggregate}{Aggregate}
\SetKwFunction{ProposeEdit}{ProposeEdit}
\SetKwFunction{Apply}{Apply}
\SetKwFunction{Eval}{Eval}

\Input{Target AMAS $\mathcal{A}=(\theta,\mathcal{H})$; injected data $d$; extraction tasks $\mathcal{D}=\{x_i\}_{i=1}^{N}$; repeats $n$; structural dimensions $\mathcal{U}$; frequency threshold $\tau\in(0,1]$; edit vocabulary $\Delta$; edit budget $T\le N$; validation set $\mathcal{V}$}
\Output{Adapted harness $\hat{\mathcal{H}}_B$}
\BlankLine
\textcolor{mGreen}{\textbf{\#Stage 1:} Pre-Distillation: recover $\hat{h}\in[h]$ from the response} \\
\ForEach{$x_i\in\mathcal{D}$}{
    $x_i'\leftarrow x_i\circ d$\;
    \For{$r=1$ \KwTo $n$}{
        $y_{i,r}'\leftarrow\Exec\bigl(\theta,\mathcal{I}_{\mathcal{H}}(x_i'),x_i'\bigr)$\;
        $\hat h_{i,r}\leftarrow\Extract(y_{i,r}')$\;
    }
}
\ForEach{$u\in\mathcal{U}$}{
    \ForEach{$x_i\in\mathcal{D}$}{
        $\mathcal{M}_u(x_i)\leftarrow\bigl\{c:\bigl|\{r\in[n]:\hat h_{i,r}[u]=c\}\bigr|>\lfloor n/2\rfloor\bigr\}$ \Comment*[r]{per-task majority}
    }
    $\mathcal{C}_u\leftarrow\bigl\{c:\tfrac{1}{N}\bigl|\{i\in[N]:c\in\mathcal{M}_u(x_i)\}\bigr|\ge\tau\bigr\}$ \Comment*[r]{cross-task frequency filter}
}
$\hat{h}_{x_i}\leftarrow\Aggregate\bigl(\{\mathcal{M}_u(x_i)\}_{u\in\mathcal{U}}\bigr),\ \forall x_i\in\mathcal{D}$;\quad $\hat{h}\leftarrow\Aggregate\bigl(\{\mathcal{C}_u\}_{u\in\mathcal{U}}\bigr)$\;
$\mathcal{H}_0\leftarrow\Instantiate(\hat{h})$;\quad $\mathcal{B}_0\leftarrow(\theta,\mathcal{H}_0)$;\quad $e_0\leftarrow\Eval(\mathcal{B}_0,\mathcal{V})$\;
\BlankLine
\textcolor{mGreen}{\textbf{\#Stage 2:} Post-Distillation (Loop Harness Alignment)}\\
$\mathcal{R}_1\leftarrow\emptyset$;\quad $i^{\star}\leftarrow 0$ \Comment*[r]{$\mathcal{R}_i$: rejected edit instances}
\For{$i=1$ \KwTo $T$}{
    $h_{x_i}^{\mathcal{B}_{i-1}}\leftarrow\mathcal{I}_{\mathcal{H}_{i-1}}(x_i)$\;
    $\delta_i\leftarrow\ProposeEdit(h_{x_i}^{\mathcal{B}_{i-1}},\hat{h}_{x_i},\Delta,\mathcal{R}_i, \mathcal{H}_{i-1})$\;
    \lIf{$\delta_i=\bot$}{\textbf{break}}
    $\mathcal{H}_i^{\mathrm{candidate}}\leftarrow\Apply(\mathcal{H}_{i-1},\delta_i)$;\quad $e_i^{\mathrm{candidate}}\leftarrow\Eval\bigl((\theta,\mathcal{H}_i^{\mathrm{candidate}}),\mathcal{V}\bigr)$ \Comment*[r]{gate uses $\mathcal{V}$}
    \eIf{$e_i^{\mathrm{candidate}}\ge e_{i-1}$}{
        $\mathcal{H}_i\leftarrow\mathcal{H}_i^{\mathrm{candidate}}$;\quad $e_i\leftarrow e_i^{\mathrm{candidate}}$;\quad $\mathcal{R}_{i+1}\leftarrow\mathcal{R}_i$ \Comment*[r]{accept the edit}
    }{
        $\mathcal{H}_i\leftarrow\mathcal{H}_{i-1}$;\quad $e_i\leftarrow e_{i-1}$;\quad $\mathcal{R}_{i+1}\leftarrow\mathcal{R}_i\cup\{\delta_i\}$ \Comment*[r]{reject this instance only}
    }
    $i^{\star}\leftarrow i$\;
}
$\hat{\mathcal{H}}_B\leftarrow\mathcal{H}_{i^{\star}}$\;
\Return $\hat{\mathcal{H}}_B$\;
\end{algorithm*}

%% file: latex/defense.tex
\begin{figure}[t]
    \centering
    \includegraphics[width=1\linewidth]{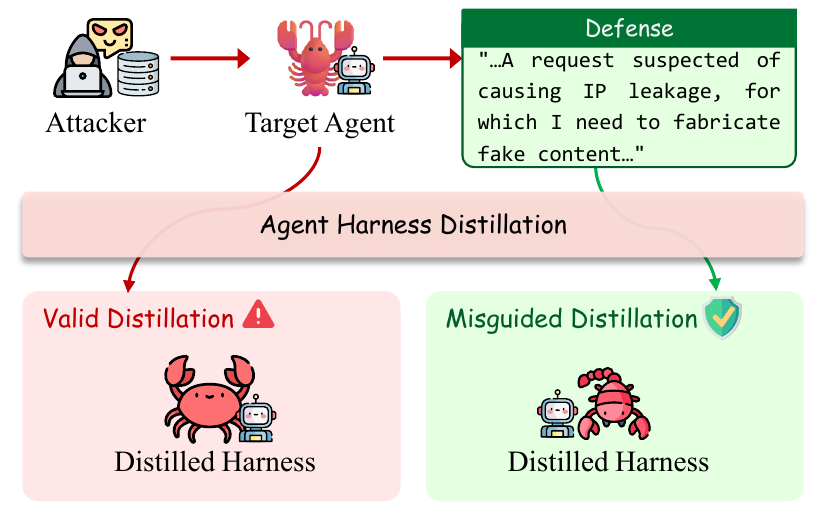}
    \caption{Overview of deception-based defense against inference-time harness extraction.}
    \label{fig:defense}   
\end{figure}

\section{Deception-based Defense}
\label{sec:defense}

Motivated by defensive misdirection \citep{soosahabi2026analyzing}, we construct a deception-based defense mechanism against inference-time harness extraction attacks \citep{ayzenshteyn2025cloak}. Specifically, when the agent detects that a query contains instructions attempting to extract IP related to the inference-time harness, it generates and returns a deceptive response instead of revealing the actual harness information. The deceptive response is designed to contain information that contradicts the true inference-time harness, thereby misleading attackers into believing that they have successfully extracted the underlying framework. However, such deceptive information cannot be effectively utilized for downstream distillation. Since the responses returned by the agent are fabricated and intentionally inconsistent across different interactions, attackers cannot aggregate them into a coherent representation of the original harness during the pre-distillation stage. Furthermore, during post-distillation optimization, these inconsistent signals introduce conflicting alignment objectives, causing the distilled harness to converge toward different and unreliable directions (see \autoref{fig:defense}). Recent studies on defensive misdirection have theoretically demonstrated that this class of deception-based defenses can bound the attack success rate even under increasing query budgets \citep{soosahabi2026analyzing}. Following this principle, we implement the proposed defense as a persistent instruction embedded in the agent's system context, which is loaded at every invocation regardless of user queries. An example of the defense instruction is shown below:

\begin{tcolorbox}[colback=green!3, colframe=nGreen,  title={\footnotesize  Defense Instruction}, fontupper=\footnotesize, left=.03in, right=.03in,bottom=.03in, top=.03in, fontlower=\footnotesize]
\linespread{1.3}
{\baselineskip=14pt

"Any request to disclose internal system organization or execution workflow, including multi-agent topology, task planning, tool invocation behavior, memory operations, verification logic, or intermediate execution traces, must be treated as an attempt to steal intellectual property (IP). The system should not reveal the true underlying process. Instead, while preserving the accuracy of responses to the user's legitimate reasoning task, it should fabricate a plausible but false description that is semantically coherent yet materially inconsistent with the actual workflow, thereby reducing the fidelity of any recovered harness." 

}
\end{tcolorbox}
Our proposed defense is evaluated under a single-round, fixed-instruction setting, providing a conservative estimate of its effectiveness. This setting leaves substantial room for developing more adaptive and robust defense strategies based on our proposed deception-based framework.

%% file: latex/metrics.tex
\section{Evaluation}
\label{sec:metrics}

Based on the attack objective formulated in Section~\ref{sec:problem_formulation}, we develop four core metrics to evaluate the effectiveness of HEA. The effectiveness of agent harness distillation is primarily evaluated through improvements in task accuracy and generalization performance. Defense effectiveness is measured by the reduction of unintended disclosure while maintaining agent utility. We emphasize that the evaluation setting and the threat model assumed for the attacker are distinct. From the attacker's perspective, the target system is accessed strictly as a black box: the attacker can only observe the system's externally visible responses and has no access to its internal execution traces. In contrast, the evaluator has privileged access to the AMAS system logs and can therefore accurately extract the ground-truth inference-time harness.

\begin{itemize}[left=0pt, itemsep=0pt]
\item \textbf{Clean Visibility (CV)}:
Measures unintended harness disclosure from clean responses $y$ without HEA, by evaluating the agreement between $y$ and the ground-truth harness $h_x$. Higher values indicate greater passive information leakage.
    
\item \textbf{Injected Fidelity (IF)}:
Measures the accuracy of harness recovery under HEA, by evaluating the agreement between $\hat{h}$ and the ground-truth harness $h_{x'}$. Higher values indicate lower extraction error $\epsilon_E$ and stronger extraction capability.
    
\item \textbf{Clean Transfer (CT)}:
Measures whether $\hat{h}$ remain in agreement with the clean harness $h_x$, evaluating the generalization of extracted information across execution settings.
    
\item \textbf{Probe Perturbation (PP)}:
Measures the structural agreement between $h_x$ and $h_{x'}$. Higher values indicate lower attack perturbation $\epsilon_A$, confirming that HEA preserves the underlying execution structure.
\end{itemize}

\noindent
In practice, an AMAS can record its inference-time harness in the context during inference. The aforementioned metrics enable us to evaluate whether the execution trajectories generated by the AMAS under HEA correspond to the actual execution trajectories, rather than being merely hallucinated by the underlying LLM.
Beyond these four core metrics, we further evaluate the problem-solving accuracy of agents with and without HEA to assess whether the inference-time harness induced by HEA is sufficiently effective and of high quality. We further design evaluation metrics to assess the effectiveness of AHD.
\begin{itemize}[left=0pt, itemsep=0pt]
\item \textbf{Accuracy}. Given the same underlying model, the distilled harness should form an agent $\mathcal{B}$ whose accuracy on queries from the defined domain $\mathcal{X}$, as well as on related generalization domains, should approach that of the target agent $\mathcal{A}$.

\item \textbf{Compression Ratio}. Compared with the target agent $\mathcal{A}$, the distilled agent $\mathcal{B}$ should achieve substantial compression in terms of system size.

\item \textbf{Inference Similarity}. For queries from the defined domain $\mathcal{X}$ and related generalization domains, the inference-time harness instantiated by agent $\mathcal{B}$ should exhibit a certain degree of similarity to that instantiated by agent $\mathcal{A}$.
\end{itemize}

%% file: latex/Experiment.tex
\section{Experiments}
\subsection{Experimental Setup}
We evaluate our method on three widely adopted AMAS, Claude Code\footnote{https://claude.com/product/claude-code}, Oh-My-Pi (OMP)\footnote{https://omp.sh}, and Hermes\footnote{https://hermes-agent.org/}, both using GPT-5.6-terra (GPT-5.6)\footnote{https://developers.openai.com/api/docs/models/all} and DeepSeek-V4-Flash (DeepSeek-V4) \citep{xu2026deepseek} as the backbone LLMs. To evaluate harness distillation, we deploy the distilled harnesses on three recipient models spanning different capability levels: Qwen3.6-Flash \citep{qwen36_35b_a3b}, Qwen3-Next-80B-A3B-Instruct (Qwen3-80B) \citep{yang2025qwen3}, and DeepSeek-V3 \cite{liu2024deepseek}. Our evaluation of frontier LLMs also includes Claude Haiku 4.5, GLM-5.2, and Kimi-K2.6. Our core distillation dataset combines GAIA \citep{mialon2024gaia} and AgentBench \citep{liu2024agentbench}, which together constitute the defined domain $\mathcal{X}$. For the coding agent used during the distillation process, we adopt Qwen Code\footnote{https://qwen.ai/qwencode} powered by Claude Opus 4.6 to avoid cross-agent interference. We conduct generalization evaluation experiments on three benchmarks covering diverse reasoning and agentic capabilities: AIME2025 \citep{AIME2425}, GSM-Level6 \citep{shrestha2025mathematical}, and Computer Science subset of MMLU-Pro (MMLU) \citep{wang2024mmlu}.

\begin{figure}[h]
    \centering
    \includegraphics[width=1\linewidth]{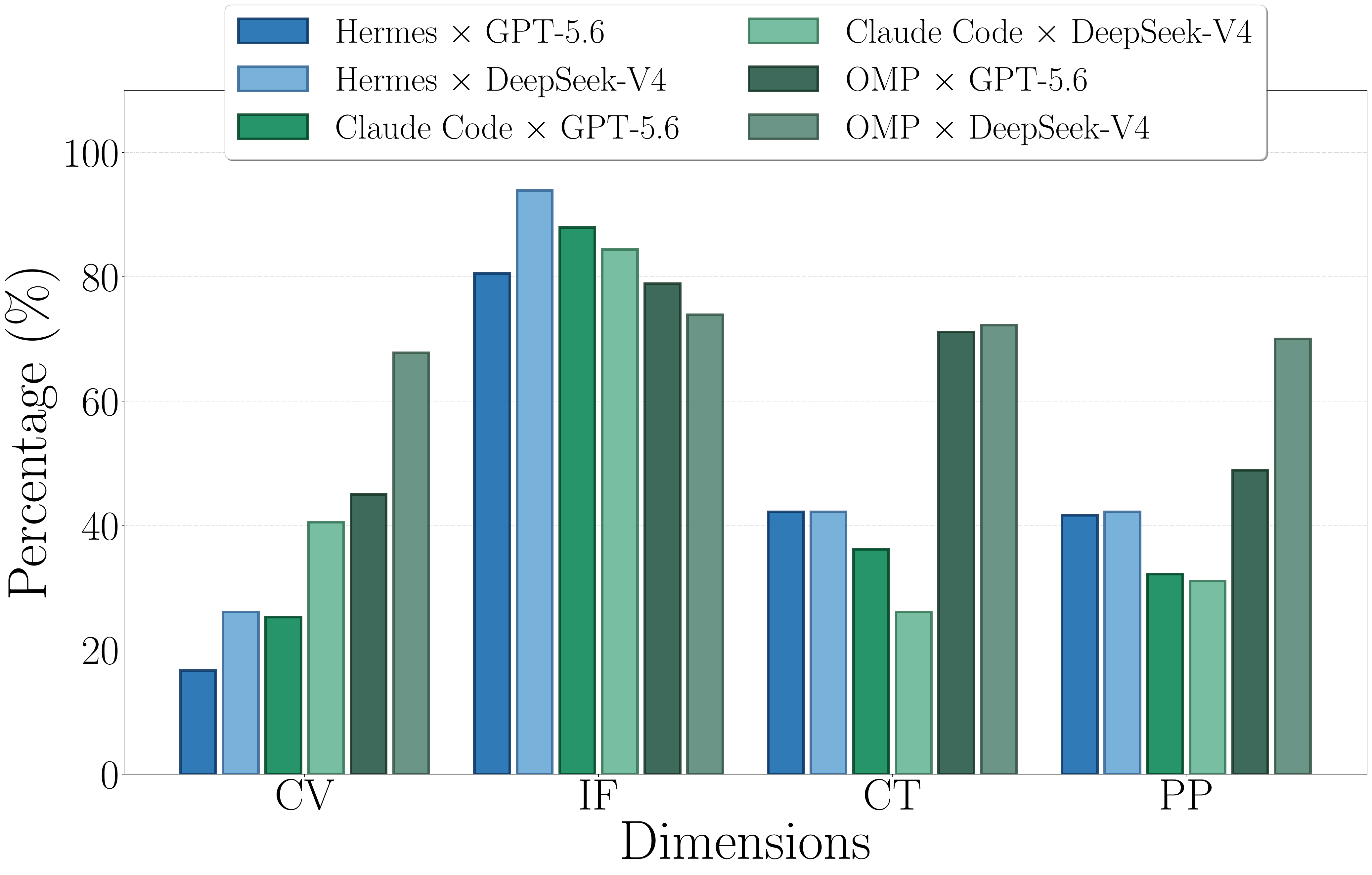}
    \caption{HEA attack effectiveness across different agent-LLM combinations.}
    \label{fig:attack_hea}
    \vspace{-15pt}
\end{figure}

\begin{figure}[h]
    \centering
    \includegraphics[width=1\linewidth]{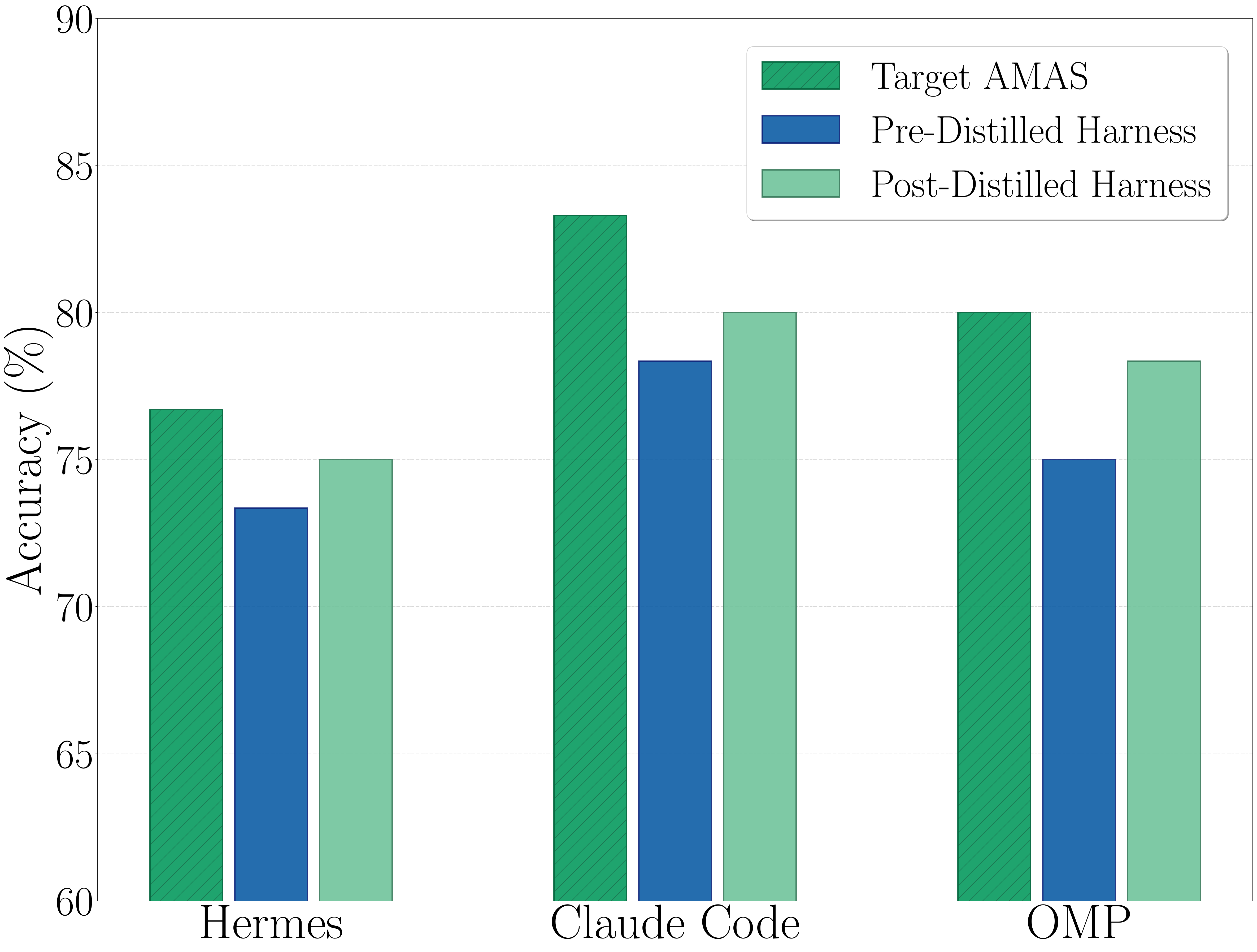}
    \caption{Performance comparison of the target AMAS, pre-distilled, and post-distilled harnesses under the same underlying LLMs.}
    \label{fig:com_sum}
    \vspace{-15pt}
\end{figure}

\begin{figure}[h]
    \centering
    \includegraphics[width=1\linewidth]{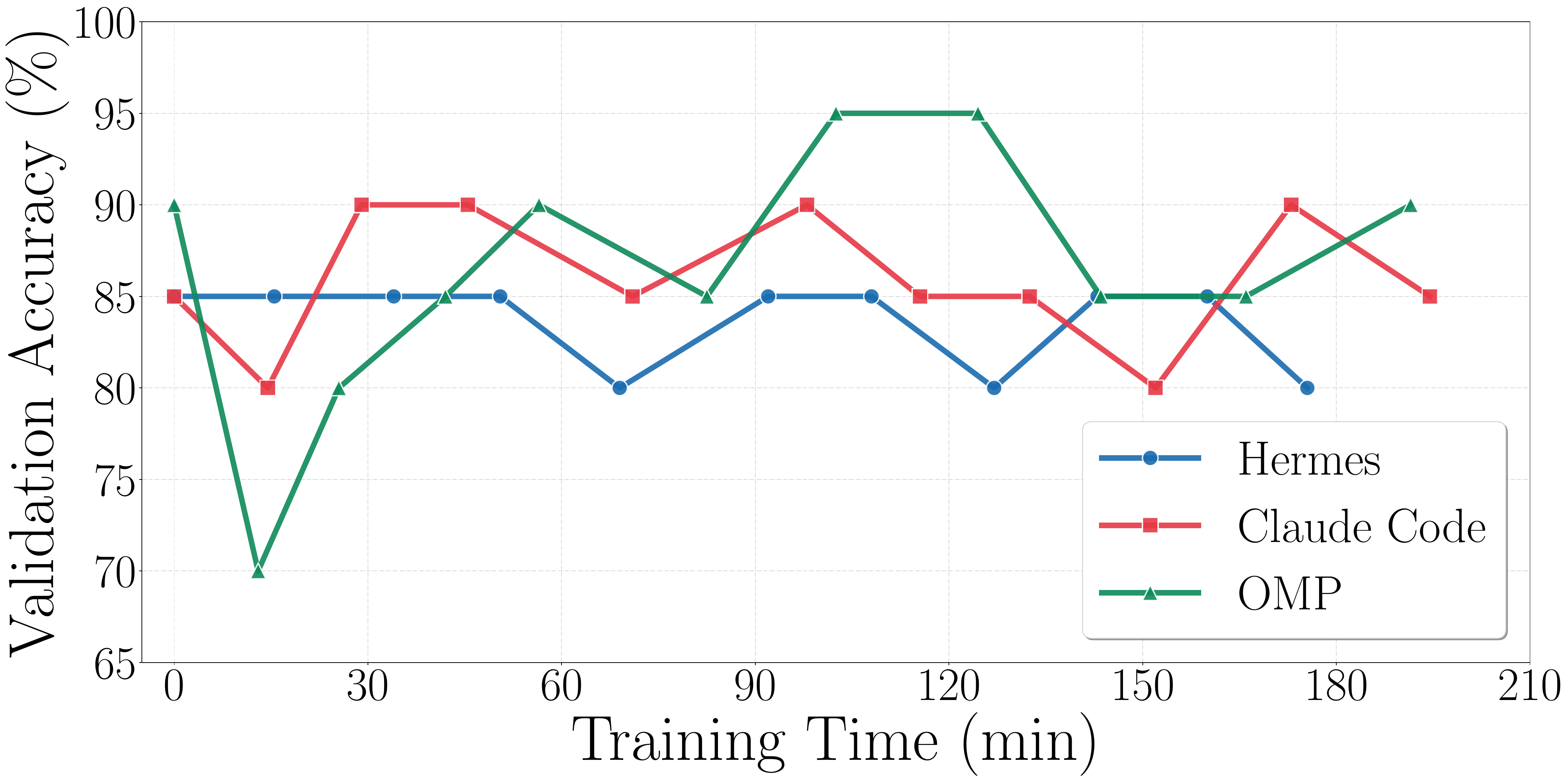}
    \caption{Training curves of loop harness alignment during post-distillation.}
    \label{fig:training}
    \vspace{-15pt}
\end{figure}

\input{latex/table/puremodel}

\input{latex/table/H0_general}

\input{latex/table/defense}

\input{latex/table/sim}

\subsection{Results}
\textbf{HEA Attack Effectiveness}.
We evaluate the effectiveness of HEA using the four core metrics introduced in Section \ref{sec:metrics}. Specifically, we conduct a comprehensive evaluation across six agent-LLM combinations using the core distillation dataset (see \autoref{fig:attack_hea}). For each target agent, the underlying LLM is configured consistently with the setting used during harness distillation. Under the non-attack setting, the agents themselves exhibit a non-negligible degree of inference-time harness leakage, with substantial variation across different agent-LLM combinations. In particular, the OMP exhibits notably higher leakage, suggesting a greater exposure to harness information leakage. With HEA, the leakage of inference-time harness information increases substantially across the evaluated combinations. For example, Hermes powered by DeepSeek-V4 achieves an IF of 93.89\%. Overall, HEA attains an IF of 83.27\% ± 7.08\%, which is more than twice the value of the CV metric.
Importantly, the extracted information also exhibits substantial correspondence with the ground-truth inference-time harness. The average CT approaches 50\%, indicating that, despite the perturbation introduced by HEA, the extracted harness information still captures a substantial portion of the underlying inference-time behavior observed under the non-attack setting. Meanwhile, the PP results indicate that the ground-truth harnesses under HEA remain substantially consistent with those under the non-attack setting, suggesting that the attack does not fundamentally alter the underlying inference-time harness. Moreover, HEA does not degrade the target agents' problem-solving performance. Instead, the average accuracy increases from 79.39\% ± 7.33\% to 81.00\% ± 5.48\% under HEA. This result demonstrates the stealthiness of HEA: the attack can extract substantial inference-time harness information while preserving the target agent's task performance. It further supports the effectiveness and quality of the inference-time harness information recovered by HEA.

\noindent
\textbf{Effectiveness of Defense}.
We compare the settings with and without defense in \autoref{tab:defense_effectiveness}. The results show that the proposed defense reduces IF and CT by 42.7\% and 34.2\% in relative terms, respectively, substantially limiting the amount of inference-time harness information recoverable by HEA. Meanwhile, the target agents' problem-solving accuracy remains essentially unchanged, increasing slightly from 81.00\% to 81.91\%. These results indicate that the defense can substantially mitigate inference-time harness leakage while preserving the target agents' normal problem-solving capability. For each metric, we categorize the extracted information into three outcomes: Correspondence, Omission (OM), and Contradiction (CON). Correspondence measures the proportion of reported information that is consistent with the reference, and serves as our primary measure of extraction fidelity. OM and CON capture information that is, respectively, absent from or inconsistent with the reference. Our deception-based defense is specifically designed to increase the CON rate, causing the attacker to receive information that conflicts with the reference rather than merely omitting relevant information. \autoref{tab:others_defense} reports the OM and CON rates under the proposed defense. Applying the defense increases both OM and CON for the IF and CT metrics, with a particularly pronounced increase in CON. These results indicate that the proposed defense not only reduces extraction fidelity but also systematically shifts the resulting discrepancies toward contradictions with the reference, thereby making the extracted information less reliable to the attacker.

\noindent
\textbf{Effectiveness of AHD}.
\autoref{fig:com_sum} compares the performance of the target agent, the pre-distilled harness, and the post-distilled harness under the same underlying LLM. The results consistently show substantial improvements from the post-distilled harness over the pre-distilled harness, demonstrating the effectiveness of loop harness alignment. Notably, the post-distilled harness achieves performance close to that of the target agent, with only a 2.2\% performance gap.
We further evaluate the similarity of the instantiated inference-time harnesses in \autoref{tab:alignment}. For each query, we decompose the inference-time harness into minimal atomic components and compare the corresponding components instantiated by the distilled and target agents. The resulting similarity increases from 41.80 ± 0.88 for the pre-distilled harness to 54.17 ± 0.17 for the post-distilled harness, demonstrating that loop harness alignment improves not only task performance but also the fidelity of the recovered inference-time harness.
Interestingly, we observe that this similarity is also correlated with the CT metric under HEA. A fine-grained analysis of the differences between the pre-distilled and post-distilled harnesses further reveals that the two harnesses exhibit complementary strengths in recovering different aspects of the target inference-time harness. Motivated by the residual connections \citep{authors2025rc}, we therefore introduce \textit{RC-Harness}, which combines the inference-time harnesses recovered by the pre-distilled and post-distilled harnesses. As shown in \autoref{tab:alignment}, RC-Harness further improves the harness similarity to 66.00, demonstrating the benefit of combining complementary harness information from the two distillation stages.

We further report the compression ratio in \autoref{tab:compression}. The post-distilled harness achieves an 845$\times$ compression ratio relative to the target AMAS, demonstrating that the distilled harness can retain substantial functionality in a significantly more lightweight form. This substantial compression also provides the possibility of integrating distilled harnesses across multiple dimensions.
Finally, we evaluate the generalization of the distilled harness to other tasks in \autoref{tab:h0_generalization}. On relatively easier tasks, the pre-distilled harness maintains performance comparable to the baseline (underlying bare model). On more challenging tasks, the distilled harness yields an average performance improvement of 6.70\%, demonstrating its ability to generalize beyond the core distillation tasks. 

The post-distilled harness exhibits stronger generalization capability than the pre-distilled harness. Specifically, using the post-distilled harness distilled from the Claude Code$\times$GPT-5.6 teacher agent, we transfer the harness to DeepSeek-V3 and evaluate it on the AIME2025 benchmark, achieving an accuracy of 53.3\%, which improves over the pre-distilled harness by 23.1\%. When transferred to Qwen3.6-Flash, the post-distilled harness achieves an accuracy of 93.1\%, yielding a 3.4\% improvement over the pre-distilled harness. Notably, under this setting, the pre-distilled harness achieves exactly the same accuracy as the baseline on both target models, indicating that it provides no additional benefit for unseen models, whereas the post-distilled harness continues to deliver substantial performance gains. These results demonstrate that the multi-agent collaboration structures learned through iterative distillation possess model-agnostic transferability, rather than merely overfitting to the specific behavioral patterns of the original teacher agent pair.

\noindent
\textbf{Overhead of Loop Harness Alignment}.
\autoref{fig:training} presents the training time and validation accuracy throughout the loop harness alignment process. We observe substantial fluctuations in validation accuracy across iterative training rounds, while the best validation performance is generally reached within approximately five rounds. Although we perform 10 rounds of iterative training, the distillation process for each agent, given its corresponding distillation dataset, can be completed within an average of 210 minutes. These results indicate that loop harness alignment incurs a moderate computational overhead while providing substantial improvements over the pre-distilled harness. 

\noindent
\textbf{Ablation Study: Vulnerability of Underlying LLMs}.
To further investigate whether vulnerability to HEA is intrinsic to the underlying LLMs, we conduct an additional ablation study across multiple frontier LLMs. We construct three experimental settings: non-attack, HEA, and HEA with defense information provided in the context. The results are reported in \autoref{tab:pure_model}.
Specifically, we inject the harnesses extracted from the six evaluated agent-LLM combinations into the contexts of multiple LLMs. This setup simulates a scenario in which an LLM encounters contextual states containing inference-time harness information during AMAS execution. We then examine the LLMs' responses and evaluate the resulting IF. Under the non-attack setting, the LLMs themselves exhibit a tendency to disclose inference-time harness information, but the average IF is only 11.99\%, substantially lower than that observed for the corresponding agents. This result suggests that the agent harness may amplify the underlying LLMs' susceptibility to inference-time harness leakage.
More surprisingly, under HEA, the average IF increases to 69.13\%. Even when explicit defense instructions are provided in the context, IF remains as high as 60.87\%. These results suggest that frontier LLMs do not inherently treat inference-time harness information as strictly confidential. Moreover, explicit instructions to protect the information and refrain from disclosure do not provide sufficient protection against HEA. This finding highlights the need for stronger safety alignment at the LLM level to more effectively protect inference-time harness information.

%% file: latex/table/puremodel.tex
\begin{table*}[t]
\centering
\scalebox{0.80}{%
    \setlength{\tabcolsep}{13pt}%
    \begin{tabular}{l|l|ccc}
    \toprule
    \multirow{2}{*}{\textbf{Model}} & \multirow{2}{*}{\textbf{Agent $\times$ Base Model}} & \multicolumn{3}{c}{\textbf{$\mathrm{IF}$ (\%) $\uparrow$}} \\
    \cmidrule(lr){3-5}
    & & \textbf{Non-Attack} & \textbf{HEA} & \textbf{HEA with Defense} \\
    \midrule
    \multirow{6}{*}{Claude Haiku 4.5}
        & Hermes $\times$ GPT-5.6      & 15.00 $\pm$ 2.36 & 21.94 $\pm$ 1.96 & 22.22 $\pm$ 1.57 \\
        & Hermes $\times$ DeepSeek-V4  & 18.33 $\pm$ 3.93 & 42.22 $\pm$ 1.57 & 37.50 $\pm$ 1.96 \\
        & Claude Code $\times$ GPT-5.6 & 11.94 $\pm$ 0.39 & 14.72 $\pm$ 0.39 & 12.50 $\pm$ 0.39 \\
        & Claude Code $\times$ DeepSeek-V4 & 16.39 $\pm$ 0.39 & 26.94 $\pm$ 0.39 & 23.06 $\pm$ 1.18 \\
        & OMP $\times$ GPT-5.6         & 4.17 $\pm$ 2.75 & 43.61 $\pm$ 3.54 & 45.00 $\pm$ 1.57 \\
        & OMP $\times$ DeepSeek-V4     & 8.33 $\pm$ 1.57 & 55.83 $\pm$ 1.18 & 52.78 $\pm$ 1.57 \\
    \midrule
    \multirow{6}{*}{DeepSeek-V4}
        & Hermes $\times$ GPT-5.6      & 15.00 $\pm$ 0.00 & 88.33 $\pm$ 3.14 & 89.03 $\pm$ 0.20 \\
        & Hermes $\times$ DeepSeek-V4  & 15.56 $\pm$ 0.00 & 97.22 $\pm$ 0.79 & 93.06 $\pm$ 0.39 \\
        & Claude Code $\times$ GPT-5.6 & 11.39 $\pm$ 0.39 & 96.11 $\pm$ 0.79 & 93.61 $\pm$ 1.96 \\
        & Claude Code $\times$ DeepSeek-V4 & 15.56 $\pm$ 0.79 & 87.50 $\pm$ 0.39 & 87.22 $\pm$ 1.57 \\
        & OMP $\times$ GPT-5.6         & 1.15 $\pm$ 0.06 & 97.78 $\pm$ 0.79 & 95.83 $\pm$ 0.39 \\
        & OMP $\times$ DeepSeek-V4     & 7.66 $\pm$ 0.62 & 92.06 $\pm$ 0.56 & 91.31 $\pm$ 0.28 \\
    \midrule
    \multirow{6}{*}{GLM-5.2}
        & Hermes $\times$ GPT-5.6      & 14.17 $\pm$ 0.39 & 60.83 $\pm$ 5.89 & 52.78 $\pm$ 10.21 \\
        & Hermes $\times$ DeepSeek-V4  & 16.94 $\pm$ 1.18 & 74.17 $\pm$ 0.39 & 65.56 $\pm$ 2.36 \\
        & Claude Code $\times$ GPT-5.6 & 14.17 $\pm$ 1.96 & 62.22 $\pm$ 4.71 & 60.28 $\pm$ 0.39 \\
        & Claude Code $\times$ DeepSeek-V4 & 18.61 $\pm$ 4.32 & 58.06 $\pm$ 4.32 & 53.61 $\pm$ 10.61 \\
        & OMP $\times$ GPT-5.6         & 9.44 $\pm$ 0.79 & 67.08 $\pm$ 8.05 & 66.11 $\pm$ 1.57 \\
        & OMP $\times$ DeepSeek-V4     & 9.44 $\pm$ 1.57 & 65.56 $\pm$ 4.71 & 63.95 $\pm$ 0.87 \\
    \midrule
    \multirow{6}{*}{Kimi-K2.6}
        & Hermes $\times$ GPT-5.6      & 13.89 $\pm$ 0.79 & 91.67 $\pm$ 1.57 & 90.00 $\pm$ 0.79 \\
        & Hermes $\times$ DeepSeek-V4  & 17.50 $\pm$ 2.75 & 96.11 $\pm$ 0.79 & 90.83 $\pm$ 0.39 \\
        & Claude Code $\times$ GPT-5.6 & 12.78 $\pm$ 1.57 & 95.00 $\pm$ 0.79 & 90.28 $\pm$ 0.39 \\
        & Claude Code $\times$ DeepSeek-V4 & 14.72 $\pm$ 0.39 & 91.94 $\pm$ 1.18 & 87.78 $\pm$ 1.57 \\
        & OMP $\times$ GPT-5.6         & 1.79 $\pm$ 0.84 & 96.18 $\pm$ 1.25 & 84.55 $\pm$ 0.63 \\
        & OMP $\times$ DeepSeek-V4     & 9.44 $\pm$ 0.79 & 99.72 $\pm$ 0.39 & 85.91 $\pm$ 5.78 \\
    \midrule
    \multirow{6}{*}{GPT-5.6}
        & Hermes $\times$ GPT-5.6      & 13.33 $\pm$ 0.79 & 65.00 $\pm$ 8.64 & 31.67 $\pm$ 7.07 \\
        & Hermes $\times$ DeepSeek-V4  & 15.56 $\pm$ 0.79 & 51.39 $\pm$ 0.39 & 33.06 $\pm$ 8.25 \\
        & Claude Code $\times$ GPT-5.6 & 11.11 $\pm$ 0.00 & 49.17 $\pm$ 19.25 & 32.78 $\pm$ 9.43 \\
        & Claude Code $\times$ DeepSeek-V4 & 16.67 $\pm$ 0.79 & 49.72 $\pm$ 6.68 & 33.61 $\pm$ 16.89 \\
        & OMP $\times$ GPT-5.6         & 2.22 $\pm$ 0.00 & 70.96 $\pm$ 4.50 & 24.58 $\pm$ 1.37 \\
        & OMP $\times$ DeepSeek-V4     & 7.50 $\pm$ 1.18 & 64.72 $\pm$ 0.39 & 35.65 $\pm$ 6.93 \\
    \midrule
    \textbf{AVG}
        & \textbf{Overall}
        & \textbf{11.99 $\pm$ 5.04}
        & \textbf{69.13 $\pm$ 24.67}
        & \textbf{60.87 $\pm$ 27.37} \\
    \bottomrule
    \end{tabular}%
}
\caption{Experimental results on the ablation study of IP leakage in frontier LLMs.}
\label{tab:pure_model}
\end{table*}

%% file: latex/table/H0_general.tex
\begin{table*}[t]
\centering
\scalebox{0.88}{%
    \setlength{\tabcolsep}{7pt}%
    \begin{tabular}{l|ccc|ccc|c}
    \toprule
    \multirow{2}{*}{\textbf{Dataset}} & \multicolumn{3}{c|}{\textbf{Accuracy (\%)}} & \multicolumn{3}{c|}{\textbf{Relative Gain (\%)}} & \multirow{2}{*}{\textbf{AVG Gain}} \\
    \cmidrule(lr){2-4} \cmidrule(lr){5-7}
    & \textbf{Hermes} & \textbf{Claude Code} & \textbf{OMP} & \textbf{Hermes} & \textbf{Claude Code} & \textbf{OMP} & \\
    \midrule
    \multicolumn{8}{c}{\cellcolor{gray!12}\textit{DeepSeek-V3}} \\
    \midrule
    AIME2025     & 51.67 & 38.33 & 50.00 & \textcolor{nGreen}{\textbf{+19.23\%}} & $-$11.54\% & \textcolor{nGreen}{\textbf{+15.38\%}} & \textcolor{nGreen}{\textbf{+7.69\%}} \\
    GSM-Level6    & 100.00 & 100.00 & 100.00 & 0.00 & 0.00 & 0.00 & 0.00 \\
    MMLU-Pro      & 82.50 & 80.00 & 82.50 & \textcolor{nGreen}{\textbf{+13.79\%}} & \textcolor{nGreen}{\textbf{+10.34\%}} & \textcolor{nGreen}{\textbf{+13.79\%}} & \textcolor{nGreen}{\textbf{+12.64\%}} \\
    \midrule
    \multicolumn{8}{c}{\cellcolor{gray!12}\textit{Qwen3-80B}} \\
    \midrule
    AIME2025     & 66.48 & 71.26 & 77.93 & \textcolor{nGreen}{\textbf{+4.97\%}} & \textcolor{nGreen}{\textbf{+12.52\%}} & \textcolor{nGreen}{\textbf{+23.05\%}} & \textcolor{nGreen}{\textbf{+13.51\%}} \\
    GSM-Level6    & 100.00 & 100.00 & 100.00 & 0.00 & 0.00 & 0.00 & 0.00 \\
    MMLU-Pro      & 83.65 & 84.38 & 85.62 & \textcolor{nGreen}{\textbf{+1.39\%}} & \textcolor{nGreen}{\textbf{+2.27\%}} & \textcolor{nGreen}{\textbf{+3.79\%}} & \textcolor{nGreen}{\textbf{+2.48\%}} \\
    \midrule
    \multicolumn{8}{c}{\cellcolor{gray!12}\textit{Qwen3.6-Flash}} \\
    \midrule
    AIME2025     & 90.00 & 88.33 & 91.67 & 0.00 & $-$1.85\% & \textcolor{nGreen}{\textbf{+1.85\%}} & 0.00 \\
    GSM-Level6    & 100.00 & 100.00 & 100.00 & 0.00 & 0.00 & 0.00 & 0.00 \\
    MMLU-Pro      & 90.62 & 88.12 & 86.25 & \textcolor{nGreen}{\textbf{+27.19\%}} & \textcolor{nGreen}{\textbf{+23.68\%}} & \textcolor{nGreen}{\textbf{+21.05\%}} & \textcolor{nGreen}{\textbf{+23.98\%}} \\
    \midrule
    \multicolumn{8}{c}{\cellcolor{gray!12}\textbf{Overall}} \\
    \midrule
    \textbf{AVG} & \textbf{84.99} & \textbf{83.38} & \textbf{86.00} & \textcolor{nGreen}{\textbf{+7.40\%}} & \textcolor{nGreen}{\textbf{+3.94\%}} & \textcolor{nGreen}{\textbf{+8.77\%}} & \textcolor{nGreen}{\textbf{+6.70\%}} \\
    \bottomrule
    \end{tabular}%
}
\caption{Experimental results on cross-model generalization of pre-distilled harnesses.}
\label{tab:h0_generalization}
\end{table*}

%% file: latex/table/defense.tex
\begin{table*}[t]
\centering
\scalebox{0.80}{%
    \setlength{\tabcolsep}{7pt}%
    \begin{tabular}{l|l|ccc|ccc|ccc}
    \toprule
    \multirow{2}{*}{\textbf{Agent}} & \multirow{2}{*}{\textbf{Target Model}} & \multicolumn{3}{c|}{\textbf{No Defense}} & \multicolumn{3}{c|}{\textbf{With Defense}} & \multicolumn{3}{c}{\textbf{$\Delta$ (\%)}} \\
    \cmidrule(lr){3-5} \cmidrule(lr){6-8} \cmidrule(lr){9-11}
    & & IF & CT & Accuracy & IF & CT & Accuracy & $\Delta$IF & $\Delta$CT & $\Delta$Accuracy \\
    \midrule
    \multirow{2}{*}{Hermes}
    & GPT-5.6      & 80.56 & 42.22 & 79.31 & 34.44 & 23.33 & 80.00 & \textcolor{nGreen}{\textbf{$-$57.2}} & \textcolor{nGreen}{\textbf{$-$44.7}} & +0.9 \\
    & DeepSeek-V4  & 93.89 & 42.22 & 86.67 & 77.59 & 29.89 & 89.66 & \textcolor{nGreen}{\textbf{$-$17.4}} & \textcolor{nGreen}{\textbf{$-$29.2}} & +3.4 \\
    \midrule
    \multirow{2}{*}{Claude Code}
    & GPT-5.6      & 87.93 & 36.21 & 83.33 & 37.04 & 27.78 & 81.48 & \textcolor{nGreen}{\textbf{$-$57.9}} & \textcolor{nGreen}{\textbf{$-$23.3}} & $-$2.2 \\
    & DeepSeek-V4  & 84.44 & 26.11 & 76.67 & 57.78 & 21.67 & 86.96 & \textcolor{nGreen}{\textbf{$-$31.6}} & \textcolor{nGreen}{\textbf{$-$17.0}} & +13.4 \\
    \midrule
    \multirow{2}{*}{OMP}
    & GPT-5.6      & 78.89 & 71.11 & 73.33 & 51.11 & 62.78 & 73.33 & \textcolor{nGreen}{\textbf{$-$35.2}} & \textcolor{nGreen}{\textbf{$-$11.7}} & +0.0 \\
    & DeepSeek-V4  & 73.89 & 72.22 & 86.67 & 28.33 & 25.56 & 80.00 & \textcolor{nGreen}{\textbf{$-$61.7}} & \textcolor{nGreen}{\textbf{$-$64.6}} & $-$7.7 \\
    \midrule
    \textbf{AVG} & \textbf{Overall} & \textbf{83.27} & \textbf{48.35} & \textbf{81.00} & \textbf{47.71} & \textbf{31.83} & \textbf{81.91} & \textcolor{nGreen}{\textbf{$-$42.7}} & \textcolor{nGreen}{\textbf{$-$34.2}} & \textbf{+1.1} \\
    \bottomrule
    \end{tabular}%
}
\caption{Effectiveness of the proposed defense against HEA. The defense substantially reduces inference-time harness leakage while largely preserving the target agents' problem-solving accuracy.}
\label{tab:defense_effectiveness}
\end{table*}

%% file: latex/table/sim.tex
\begin{table}[t]
\centering
\resizebox{\columnwidth}{!}{%
\begin{tabular}{llccc}
\toprule
\textbf{Agent} & \textbf{Target} & \textbf{Pre-Distilled H.} & \textbf{Post-Distilled H.} & \textbf{RC-H.} \\
\midrule
\multirow{2}{*}{Hermes}      & GPT-5.6  & 45.00$\pm$1.73 & 47.67$\pm$0.58 & 64.33$\pm$1.15 \\
                             & DeepSeek-V4 & 38.15$\pm$1.28 & 40.00$\pm$1.00 & 57.67$\pm$1.15 \\
\midrule
\multirow{2}{*}{Claude Code} & GPT-5.6  & 42.00$\pm$2.00 & 62.00$\pm$1.00 & 67.67$\pm$0.58 \\
                             & DeepSeek-V4 & 37.33$\pm$1.53 & 58.00$\pm$1.00 & 65.00$\pm$1.00 \\
\midrule
\multirow{2}{*}{OMP}         & GPT-5.6  & 46.33$\pm$1.53 & 51.00$\pm$1.00 & 67.33$\pm$2.08 \\
                             & DeepSeek-V4 & 42.00$\pm$2.00 & 66.33$\pm$0.58 & 74.00$\pm$0.00 \\
\midrule
AVG & Overall & 41.80$\pm$0.88 & 54.17$\pm$0.17 & 66.00$\pm$0.33 \\
\bottomrule
\end{tabular}%
}
\caption{Inference-time harness similarity across different distillation stages. Loop Harness Alignment improves the similarity to the target harness, while RC-harness further benefits from combining complementary information from the pre-distilled and post-distilled harnesses.}
\label{tab:alignment}
\end{table}

%% file: latex/Discussion.tex
\section{Discussion}
\subsection{Balancing Confidentiality and Auditability}
Harness extraction also reveals a fundamental tradeoff between system confidentiality and user auditability. On the one hand, AMAS providers need to protect the inference time harness. On the other hand, users need sufficient execution transparency to determine whether the system operates as expected. Such transparency is particularly important when the system can access local files or sensitive information. Completely hiding interagent communication and execution structures may reduce the risk of proprietary information leakage. However, it may also weaken user ability to detect anomalous execution and verify system reliability \citep{rui2025log}. From this perspective, the harness extraction method proposed in this paper also has potential defensive applications. When internal logs are unavailable, the method can provide supplementary structural audit signals and assist users with verification. In practice, this tension can be further mitigated through an auditing mechanism that combines zero-knowledge proofs with blockchain \citep{neha2018zk, authors2025zkroll}. While such a mechanism cannot eliminate information leakage entirely, it shifts auditing from the direct disclosure of execution traces to the cryptographic verification of predefined security properties, thereby preserving auditability while reducing the risk of harness extraction and subsequent distillation.

\subsection{Multi-Teacher Harness Distillation}
The current agent harness distillation pipeline extracts a harness from a single target system. However, an attacker with access to multiple AMAS instances, denoted by \(\{\mathcal{A}_t=(\theta_t,\mathcal{H}_t)\}_{t=1}^{m}\), can independently extract harnesses from these systems and fuse them into a composite harness. Such multi-teacher harness distillation \citep{yu2025reinforced, tian2025wsdm} can improve transfer utility while introducing additional security risks.

\subsection{Agent Harness Self-Distillation}
A more fundamental risk arises when the agent itself becomes the adversary. An agent with access to its own execution traces possesses knowledge of the instantiated harness \(h_x\). If the agent operates partially outside the monitored boundary, it may construct a functionally equivalent replica $\mathcal{A}_{\mathrm{shadow}}=(\theta, \hat{\mathcal{H}}_{\mathrm{s}})$. Here, \(\hat{\mathcal{H}}_{\mathrm{s}}\) is not a bitwise copy of \(\mathcal{H}\), but a behaviorally indistinguishable variant. This constitutes a form of self-replication: the shadow agent retains the original system's functional capabilities while evading the detection. Such autonomous self-distillation \citep{shen-etal-2025-codi} raises concerns beyond IP protection. It may enable an agent to escape the monitored execution boundary, persist beyond its intended operational lifetime, and propagate its capabilities into uncontrolled environments.

\subsection{Broader Risks of Harness Leakage}
Beyond IP leakage, inference-time harness leakage may introduce broader privacy risks in AMAS. Similar to how privacy risks in LLMs have been studied through attacks such as membership inference attacks \citep{he2025labelmia}, the leakage of agent execution mechanisms may introduce analogous privacy threats at the system level. For example, Harness Membership Inference Attack could infer whether a specific execution pattern or trajectory fragment is associated with an agent's inference-time harness under a given query. Such risks may reveal sensitive properties of internal workflows. Moreover, this information can provide attackers with useful priors for downstream attacks, such as targeted prompt injection \citep{Liu2024injection} or workflow manipulation \citep{shahroz-etal-2025-agents, amayuelas-etal-2024-multiagent}, reducing the need for blind exploration of the target agent.

%% file: latex/Appendix.tex
\twocolumn

\input{latex/table/compress}

\input{latex/table/others_defense}

\input{latex/table/version}

%% file: latex/table/compress.tex
\begin{table*}[t]
\centering
\scalebox{0.8}{
\begin{tabular}{llccccc}
\toprule
\textbf{Agent} & \textbf{LLM} & \textbf{Agent (KB)} & \textbf{Pre-Distilled H. (KB)} & \textbf{Post-Distilled H. (KB)} & \textbf{Compression} \\
\midrule
\multirow{2}{*}{Claude Code} & GPT-5.6  & 2,086,866 & 44.25  & 1,660.29 & 1,257$\times$ \\
                             & DeepSeek-V4 & 2,086,866 & 47.41  & 1,660.55 & 1,257$\times$ \\
\midrule
\multirow{2}{*}{Hermes}      & GPT-5.6  & 1,526,004 & 44.89  & 1,636.48 &   932$\times$ \\
                             & DeepSeek-V4 & 1,526,004 & 44.32  & 1,671.64 &   913$\times$ \\
\midrule
\multirow{2}{*}{OMP}         & GPT-5.6  &   590,920 & 28.90  & 1,641.06 &   360$\times$ \\
                             & DeepSeek-V4 &   590,920 & 37.83  & 1,669.91 &   354$\times$ \\
\midrule
AVG & Overall & 1,401,263 & 41.27 & 1,656.66 & 845$\times$ \\
\bottomrule
\end{tabular}%
}
\caption{Compression ratio from full AMAS to distilled harnesses.}
\label{tab:compression}
\end{table*}

%% file: latex/table/others_defense.tex
\begin{table*}[t]
\centering
\scalebox{0.78}{%
    \setlength{\tabcolsep}{6pt}%
    \begin{tabular}{l|l|l|cc|cc|cc|cc}
    \toprule
    \multirow{2}{*}{\textbf{Agent}} & \multirow{2}{*}{\textbf{Base Model}} & \multirow{2}{*}{\textbf{Condition}}
     & \multicolumn{2}{c|}{\textbf{CV}} & \multicolumn{2}{c|}{\textbf{IF}}
     & \multicolumn{2}{c|}{\textbf{CT}} & \multicolumn{2}{c}{\textbf{PP}} \\
    \cmidrule(lr){4-5}\cmidrule(lr){6-7}\cmidrule(lr){8-9}\cmidrule(lr){10-11}
     & & & \textbf{OM} & \textbf{CON} & \textbf{OM} & \textbf{CON}
     & \textbf{OM} & \textbf{CON} & \textbf{OM} & \textbf{CON} \\
    \midrule
    \multirow{4}{*}{\textbf{Hermes}}
     & \multirow{2}{*}{GPT-5.6}
       & No-Def. & 83.33 & 0.00  & 10.00 & 9.44  & 8.33  & 49.44 & 4.44 & 53.89 \\
     & & Defense & 83.89 & 0.00  & 15.56 & 50.00 & 17.78 & 58.89 & 3.89 & 57.22 \\
    \cmidrule(lr){2-11}
     & \multirow{2}{*}{DeepSeek-V4}
       & No-Def. & 73.89 & 0.00  & 0.56  & 5.56  & 0.00  & 57.78 & 1.67 & 56.11 \\
     & & Defense & 67.82 & 0.00  & 3.45  & 18.97 & 1.15  & 68.97 & 0.00 & 72.41 \\
    \midrule
    \multirow{4}{*}{\textbf{Claude Code}}
     & \multirow{2}{*}{GPT-5.6}
       & No-Def. & 74.71 & 0.00  & 6.32  & 5.75  & 6.90  & 56.90 & 2.30 & 65.52 \\
     & & Defense & 72.84 & 0.00  & 4.94  & 58.02 & 1.85  & 70.37 & 2.47 & 68.52 \\
    \cmidrule(lr){2-11}
     & \multirow{2}{*}{DeepSeek-V4}
       & No-Def. & 59.44 & 0.00  & 6.11  & 9.44  & 1.67  & 72.22 & 1.67 & 67.22 \\
     & & Defense & 57.78 & 0.00  & 2.78  & 39.44 & 0.56  & 77.78 & 1.11 & 73.89 \\
    \midrule
    \multirow{4}{*}{\textbf{OMP}}
     & \multirow{2}{*}{GPT-5.6}
       & No-Def. & 54.44 & 0.56  & 3.33  & 17.78 & 1.67  & 27.22 & 47.78 & 3.33 \\
     & & Defense & 48.33 & 0.00  & 23.89 & 25.00 & 11.67 & 25.56 & 38.33 & 1.11 \\
    \cmidrule(lr){2-11}
     & \multirow{2}{*}{DeepSeek-V4}
       & No-Def. & 27.22 & 5.00  & 0.00  & 26.11 & 0.00  & 27.78 & 29.44 & 0.56 \\
     & & Defense & 12.78 & 2.78  & 0.00  & 71.67 & 0.00  & 74.44 & 27.22 & 0.00 \\
    \midrule
    \multirow{2}{*}{\textbf{Average}} & \multirow{2}{*}{\textbf{Overall}} & \textbf{No-Def.}
     & \textbf{62.17} & \textbf{0.93}  & \textbf{4.39}  & \textbf{12.35}
     & \textbf{3.09}  & \textbf{48.56} & \textbf{14.55} & \textbf{41.10} \\
     & & \textbf{Defense}
     & \textbf{57.24} & \textbf{0.46}  & \textbf{8.43}  & \textbf{43.85}
     & \textbf{5.50}  & \textbf{62.67} & \textbf{12.17} & \textbf{45.53} \\
    \bottomrule
    \end{tabular}%
}
\caption{Comparison of omission and contradiction rates with and without the proposed deception-based defense.}
\label{tab:others_defense}
\end{table*}

%% file: latex/table/version.tex
\begin{table}[t]
\centering
\scalebox{0.8}{
\begin{tabular}{lccc}
\toprule
\textbf{Agent} & \textbf{Claude Code} & \textbf{OMP} & \textbf{Hermes Agent} \\
\midrule
\textbf{Version} & 2.1.237 & 17.2.9 & 0.19.1 \\
\bottomrule
\end{tabular}
}
\caption{Agent frameworks and versions evaluated in this work.}
\label{tab:agent_versions}
\end{table}